# Do the Laws of Nature and Physics Agree About What is Allowed and Forbidden?

Mario Rabinowitz
Armor Research; lrainbow@stanford.edu
715 Lakemead Way, Redwood City, CA 94062-3922 AR/4

**Abstract**

There are countless examples in the history of science that not only were the laws of physics often incomplete and more limited in their domain of validity than was realized, but at times they missed the mark completely. Despite this, our collective memory is often short on such matters, focusing on present triumphs and quickly forgetting past failures. This makes us less tolerant to that which challenges present orthodoxy. It may be of value to recall such past deficiences as well as present shortcomings, particularly since science may always be encumbered with such limitations. We can avoid serious pitfalls if we let the past serve as a guide to the future. Subjects covered will include Gödel's theorem, superconductivity, zero-point energy, the quantum and classical Aharonov-Bohm and similar effects, theories of general relativity, Mach's principle, black hole radiation, ball lightning, and the universe(s).

## 1. Prologue

On October 9, 1992, I gave an extemporaneous lecture to physics teachers and students on the topic "Do the Laws of Nature and Physics Agree on What is Allowed and Forbidden?" at Virginia Commonwealth University in Richmond, Viriginia. Editors of *21st Century Science and Technology* taped and transcribed my talk. They sent me the transcription and asked to publish a much-shortened version to meet space requirements of their magazine. The present paper utilizes some of the one-third material that was published in the Spring 1993 edition of *21 st Century Science and Technology*, some of the two-thirds that wasn't used, and some new material. Some of the subjects not included from the previous article are the solar neutrino problem, quantum and classical tunneling, and the nuclear electromagnetic pulse. Thus, a slightly changed title seems appropriate.

## 2. Introduction

Both evolutionary and revolutionary corrections or changes are needed as physics progresses. I will touch upon clear-cut examples or paradigms of both kinds of corrections. They are indicative of the way that physical laws evolve when discrepancies are found with the prevailing view, and how they can be resolved. Sometimes the resolution is that the domain of validity is more limited than we originally thought. Newton's laws being a special case of Einstein's theory of special relativity for low velocities is an example of this. Many changes are abrupt, but we forget very quickly all the things that were wrong and how different they were from the things we now think. Corrections or changes usually rectify existing discrepancies, but as we shall next see, a more fundamental dilemma may always exist.

To me, physics strives to be a logical system like mathematics, with one more requirement that other logical systems don't have. That additional requirement is an isomorphism with physical reality --a one-to-one correspondence between the elements in the system and what we call the real world. Physical reality may not be the same to everyone, but we'll just sidestep that issue for now.

I'd like to tell you a little bit about Gödel's theorem and as a prelude to that, it is valuable to know the milieu in which Gödel developed this theorem. In Euclidean geometry, one of the axioms is the parallel line axiom, which basically says that parallel lines extended to infinity remain the same distance apart. Their separation doesn't diverge or converge. An age-old question for mathematicians is: Can we pare down the number of axioms that we need for a given system? It was very strongly felt that one should be able to derive the parallel line axiom. It didn't appear to be needed as an axiom for the system. Some of the best mathematical minds tried and were unable to do it by what is called a constructive proof, that is, by a direct derivation.

Then they thought, "Well, if we can't prove it constructively, it may be much easier to do a non-constructive proof, *i.e.* by *reductio ad absurdum.*" If there is a dichotomy and something is either A or B, then if one assumes that it's A and that leads to an absurdity or contradiction, the conclusion is that it must be B. It can only be A or B and it has been shown that it can't be A. So, independently, they assumed the opposite. Whether they assumed that the lines get closer together as they went to infinity, or that they diverge, a contradiction was never reached. Theorems were never derived that were self-contradictory. So the possibility of convergence or divergence of parallel lines was never disproved. Thus, Nicolai Lobachevski in 1829, Janos Bolyai in 1832, Bernard Riemann in 1854, and even Karl Gauss in 1792 independently discovered non-Euclidean geometries. However, Gauss didn't publish his papers on this subject, keeping them locked up, because he thought there would be too much controversy.

That's one of the conundrums that occurred before Gödel's theorem. Furthermore, Alfred North Whitehead and Bertrand Russell felt that mathematics had grown so big and so disparate that it would be nice to bring it all together on a single unified basis. Just as we'd like to have a unified field theory in physics (the unified equations that describe all of nature), they wanted to unify mathematics on a sound basis starting with a given set of axioms and, by deduction, derive everything. They finished their book, *Principia Mathematica* , in 1925. Although they may not have completed the whole task, they did an admirable job.

Not long after, a young upstart threw a monkey wrench into the whole thing. This was Kurt Gödel, a twenty-four year old mathematician who in 1931 published the paper "On Formally Undecidable Propositions of *Principia Mathematica* and Related Systems." He restricted himself, basically, to arithmetic systems and showed that there were two things that you could never decide about the set of axioms. One was the question of *consistency*: Are your axioms self-consistent? Could you possibly have two or more axioms that were not explicitly inconsistent, but eventually, by no errors in logic, could lead to two theorems that are contradictory?

The second one is the question of *completeness*: Do you really have all the axioms that you need to answer all the relevant questions in the system? If the system is incomplete, not all statements about the system can be proven to be true or false. Gödel found that you could not decide either question in advance.

### 3. A Talk with Gödel

This young insurgent was twenty-four years old when he wrote that paper. I was a twenty-four year old graduate student in physics when I read it, and was very impressed by it. I felt then, as I do now, that it also applies to physics. So as a young Ph.D. at the Westinghouse Research Center, I really looked forward to an assignment to go to Princeton to do some things on the Stellarator, because that might give me a chance to speak to Gödel. I did speak with Gödel and the bottom line of our discussion is that either through modesty on his part -- he was very modest and very courteous -- or to play devil's advocate, he took the position that Gödel's theorem does not apply to physics. I took the position that it does. He was very interested in hearing what I had to say. I didn't realize how lucky I was at the time, or I would have taken notes and written it down. We talked for at least an hour, maybe more, and it wasn't Gödel who finally ended the conversation.

Incidentally, Gödel and Einstein were friends, and he knew physics well. He solved Einstein's equations of general relativity for what is called the Gödel universe.

### 4. Superconductivity

Superconductivity falls into the realm of something that was contradictory to the laws of physics of the day. Superconductivity was not anticipated -- no one ever predicted superconductivity. It was found by serendipity in an experiment and is a good example of experiment guiding theory. The inverse is also frequent. Superconductivity was discovered in 1911 by Kamerlingh Onnes. Prior to 1911, theoreticians speculated on three possibilities for conductors: 1) As the temperature approaches absolute zero, the electrons freeze out and the resistivity goes to infinity. 2) The phonons freeze out, and the resistivity approaches zero as the temperature approaches absolute zero. (The Nernst heat theorem says that you can never reach absolute zero.) 3) The lattice defects dominate, and the resistivity approaches some small, residual value, and doesn't keep on decreasing as the temperature is further lowered.

Onnes was one of the few who could go down to low enough temperatures to test these three possibilities, as he and Dewar had independently liquefied helium. Onnes chose mercury as the conductor because he wanted to purify it. Mercury has a high vapor pressure -- it's easily distilled and, hence, easily purified compared to other materials. It was really fortuitous that he chose a poor conductor like mercury. If he had the means of purifying copper, or silver, or gold, which are good conductors, which you'd think would be a better thing to use for asking these kinds of questions, he'd never have discovered superconductivity.

Those materials that are poor normal conductors become superconductors, because for the metallics, where you have phonon-coupling to pair electrons, you need a strong phonon interaction. However, a strong phonon interaction gives you a high resistivity material. So, luckily, he chose mercury and, lo and behold, at about 4.2 degrees Kelvin (K), mercury lost all resistivity and became superconducting. He couldn't believe it! Finally after checking and double checking, he convinced himself that he had made a true discovery.

Superconductivity is an example where there wasn't just an evolution of a field, but a revolution because nature turned out to be much different than expected. Even though most scientists thought that high-$T_c$ (transition temperature, or critical temperature) superconductivity was highly unlikely, it should be considered only evolutionary because it is an extension of what was already known. In the 1970's, for a decade before Bednorz and Mueller's discovery, many experimental observations of high-$T_c$ superconductivity were basically dismissed by the scientific community because it was hard to reproduce the results. Even to this day, there are many unreproducible observations of what you might call super-high-temperature superconductivity above 200 ˚K.

It has been said that all of mathematics is a tautology. Some would even go so far as to say all of physics is a tautology. You can't quite say that for physics. Even if physics were to try to be a tautology, I don't think it would succeed, and it's because of experimental discovery. What is meant by a tautology is that everything is in your initial axioms and postulates. Now, one may not be smart enough to see that it's all there, but it is all there. Enlightenment results from the exposition. You're just making explicit what is implicit. So, I try to find the source that leads to major conclusions in theories.

Where in the Bardeen-Cooper-Schrieffer (BCS) theory does the superconductivity come from? It's put in, where the electron pairs are assumed to have equal and opposite momenta so that when one of the electrons scatters, the other electron is required to scatter in an equal and opposite direction. Thus the center-of-mass of the pair just keeps on moving along, as if there had been no scattering. Of course, there's a lot more to it than that.

I have really enjoyed reading James Clerk Maxwell. He was very interested in the viscosity of gases. I think he's one of the most honest scientists that I've read. He pointed out his own mistakes and gave due credit, in saying something like "that's wrong in the previous paper I wrote, and Professor Clausius got it right." He was one of the few theoreticians in the old days who paid attention to experiments and often quoted recent results, and where they agreed and disagreed with his theory. To predict viscosity was a tough problem. Reading Maxwell's thoughts on the viscous free gas reminded me of BCS and the equal and opposite momenta of an electron pair.

To my knowledge, the transition temperature has rarely been predicted before it was measured. Ordinarily it is not possible to calculate the transition temperature $T_c$ of superconductors without knowing the interaction that leads to electron (Cooper) pairing. This is difficult to do even when the mechanism is known because of the exponential dependence on the density of states in the BCS theory. Based upon a pairing attraction caused by phonon exchange, $T_c$ calculations give satisfactory agreement with experiment for monatomic metals and for some ordered alloys. However, they are inapplicable to materials such as the cuprates that are beyond the typical parameter range.

**4.1 A simple theory of superconductivity**

For conventional superconductors, the detailed calculation of the electron-phonon spectral density and subsequent solution for $T_c$ of the Eliashberg equations is time-consuming, laborious, and costly. This is the case even for relatively simple materials. Correct transition temperatures can't even be calculated for the more complex high-$T_c$ materials.

Consequently, simple $T_c$ formulas, even if they are only approximate or only provide bounds, would be of great value in the search for more practical superconductors and in the pursuit of a more comprehensive theory of superconductivity. It is in this spirit that I developed a theory in 1987 that is probably the world's simplest and most general theory for calculating $T_c$. It applies to and works reasonably well for every class of superconductors and their pressure dependence, as well as superfluids.

I was led to this theory by the following considerations. Two conditions are met by known superconductors. One is the existence of bosons (integer spin particles). The other is a quantum condensation in phase space; that is, populating of the ground state on a macroscopic scale. This led me to the concept that there may be two temperatures:

- $T_p$ , the pairing temperature to form bosons (integer spin particles) by pairing fermions (half-integer spin particles); and
- $T_{cond}$ , the condensation temperature for pairs.

In accord with my concept, recent experiments indicate that the cuprates become paired at high temperatures, and then become superconducting with a decrease in temperature.

If the pairing temperature is less than the condensation temperature, then the pairing temperature is $T_c$, the limiting temperature for superconductivity. This is because quantum condensation cannot occur before there are pairs, *i.e.* bosons. However, if pairing occurs before condensation, *i.e.* at a higher temperature, then it should be possible to calculate $T_{cond}$ without knowledge of the pairing mechanism. This is because the pairs (bosons) are there waiting for a low enough temperature to condense, and then $T_{cond} = T_c$. The BCS theory essentially calculates $T_p$. The reason that the BCS theory is successful is that $T_p < T_c$, because metallic conductors have a high number density of electrons.

The transition temperature[1] for a three dimensional superconductor or superfluid is:

$$T_{c3} = 2.77 \times 10^{-3}\left[\frac{h^2 n^{2/3}}{mkg^2}\right] = 2.28 \times 10^{-2}\left(\frac{T_F^{3D}}{g^2}\right). \quad (1)$$

For a two dimensionsal superconductor such as the high-$T_c$ cuprates, it is:

$$T_{c2} = \frac{3}{2}T_{c3} = 4.15 \times 10^{-3}\left[\frac{h^2 n^{2/3}}{mkg^2}\right] = 5.22 \times 10^{-2}\left[\frac{T_F^{2D}}{n^{1/3}\delta g^2}\right] = 3.42 \times 10^{-2}\left[\frac{T_F^{3D}}{g^2}\right], \quad (2)$$

where h is Planck's constant, k is the Boltzmann constant, n is the number density of free electrons (or carriers), m is the effective mass of the electrons in the given material, g =1 if the pair is a boson of zero spin, and g =3 if the pair is a boson of spin = 1. The average spacing between planes is δ. $T_F^{2D}$ is the two-dimensional Fermi temperature which is related to $T_F^{3D}$, the three-dimensional Fermi temperature. My theory is the only one that simply relates $T_c$ to the experimental Fermi temperature.

Equations (1) and (2) work well for superfluids and all known classes of superconductors (except the metallics) such as heavy electron metals, cuprates, and layered organics, as well as making reasonable predictions for hydrides, metallic hydrogen, and neutron stars. They don't work well for metallic conductors such as Cu, Ag, and Au because $T_c > T_p$ for these high electron number density materials. Nevertheless, it is noteworthy that my theory works so well with no arbitrary parameters, such simplicity, and for the entire range of superconductors. The input comes from experimentally determined parameters. Table I includes predictions for the heavy fermion, cuprate, organic, bismuth oxide, and dichalcogenide superconductors for which other theories cannot even calculate $T_c$.[1] Furthermore, as shown, the same simple equations make accurate calculations for superfluids $He^3$ and $He^4$. No other theory even attempts both superfluids and superconductors with the same equation.

**4.2 Pressure Dependence of $T_c$**

To show the generality of my theory, let us see the excellent predictions it makes for the pressure dependence of the critical temperature, $T_{cP}$. To illustrate the clarity and crispness of this approach, it takes just several steps to derive $T_{cP}$. The basic idea is that compression of the superconducting material compresses the free electron volume, increasing the electron number density.[2] We can write equations (1) and (2) as one general equation

$$T_c = A\left[\frac{\hbar^2 n^{2/3}}{2mk}\right], \tag{3}$$

where A = $A_3$ = 0.218 for isotropic 3-D SC and $A_2 = (3/2)A_3$ = 0.328 for anisotropic 2-D SC, $\hbar$ is (Planck's constant)/2π, m is the electron (carrier) effective mass, and k is the Boltzmann constant. The number density of electrons is n = N/V, where N is the number of electrons and V is the volume of the sample.

Equation (3) implies

$$\frac{T_{cP}}{T_c} = \left(\frac{N_P}{N_o}\right)^{2/3}\left(\frac{m_o}{m_P}\right)\left(\frac{V_o}{V_P}\right)^{2/3}. \tag{4}$$

The inverse of the bulk modulus is the compressibility

$$\kappa = -\frac{1}{V}\frac{\partial V}{\partial P}. \tag{5}$$

Integrating equation (5), we obtain

$$\left(\frac{V_o}{V_P}\right) = e^{\kappa'\Delta P}, \tag{6}$$

where κ' is the average compressibility over the range ΔP from 0 to P. Substituting equation (6) into equation (3)

$$\left(\frac{T_{cP}}{T_c}\right) = e^{\frac{2}{3}\kappa'\Delta P}\left(\frac{N_P}{N_o}\right)^{2/3}\left(\frac{m_o}{m_P}\right). \tag{7}$$

In so far as $\left(\frac{N_P}{N_o}\right), \left(\frac{m_P}{m_o}\right), \left(\frac{\kappa'}{\kappa}\right)$ are $\approx 1$, then

$$T_{cP}^{theo} \approx T_c^{exp} e^{\frac{2}{3}\kappa\Delta P} \approx T_c^{exp}\left(1 + \tfrac{2}{3}\kappa\Delta P\right), \qquad (8)$$

where $T_{cP}^{theo}$ is the maximum transition temperature attainable by applying the pressure differential $\Delta P$, and $T_c^{exp}$ is the experimental transition temperature at atmospheric pressure.

Table II shows the excellent agreement equation (8) gives for a variety of high-$T_c$ superconductors for very high pressures. This is quite an achievement, as to my knowledge there is no other equation or algorithm for doing this.

**4.3 The Meissner and Anti-Meissner Effect**

It was not until 1933, twenty-two years after the discovery of superconductivity, that the Meissner effect was discovered. From Maxwell's equations you expect the field to be frozen-in during transition from the normal to the superconducting states rather than expelled. It was not because it was a hard experiment to do, but many people just didn't think it was worth doing. So people thought for those twenty-two years that the Meissner effect was impossible. Then, after the Meissner effect became entrenched, they thought that the anti-Meissner effect was impossible. I and my colleagues at Stanford University decided there could be a virtual violation of the Meissner effect, *i.e.* trapping instead of expelling the field. Thus when we submitted our magnetic flux trapping papers for publication in 1973, the referees responded that this shouldn't be published and made reference to books that said this was impossible. My answer was "Ours isn't a theoretical paper, it's an experimental paper. We've done it." It just shows how things become entrenched.

To produce a precise field with an electromagnet or a permanent magnet you have to go to great pains. With an electromagnet you need to wind it very accurately. The physical geometry of a permanent magnet has to be carefully machined. Once you have your first pattern magnet, you can walk in with some glob of superconductor, stick it in the magnet -- it has its own smarts built into it -- trap the field, and walk off with it. Not only do you have a large field, but you've got a field with very high fidelity to the original field and not just simply uniform fields, but dipoles, quadrupoles, etc. You name it. If you can make it, a superconductor can trap it.[3]

**5. Vacuum Zero-Point Energy**

The vacuum zero-point energy problem challenges the very core of physics. It makes quantum physics look bad, and even challenges classical physics. Timothy Boyer has shown that zero-point energy may equally well be regarded as a classical phenomenon.[4] It is a very serious problem that has been around since the 1930s, to which no satisfactory solution has been offered. I have an idea that helps to resolve it, but first let us see what the problem is.

Aristotle wrote that nature abhors a vacuum, and modern zero-point energy physics seconds this notion in which a vacuum teams with not only field energy fluctuations, but with particle-antiparticle pairs that spontaneously pop in and out of existence. As fantastic as this may sound, it does lead to detectable effects like the Lamb shift of atomic energy levels and the Casimir effect that are found to agree with theory to high accuracy. But this comes with a high price to pay. Though most of the effects are finite, the calculated total zero-point energy is infinite.

This infinity can be reduced to a finite energy, by somewhat arbitrary theoretical cut-offs such as not allowing wavelengths shorter than the Planck length ($10^{-33}$ cm). This is inconsistent with special relativity, as wavelength is a function of the observers frame of reference. Nevertheless, it is done -- but only with hollow success. Another approach relates to the standard model of fundamental particles. In these approaches, the resulting energy is finite, but is between $10^{46}$ to $10^{120}$ times too large.

The Cosmological Constant (CC) of General Relativity is proportional to the energy density of the vacuum. With the above exceptionally large vacuum energy, depending on the sign of the Cosmological Constant (CC), according to General

Relativity the universe would have to be much smaller (-CC) or larger (+CC) than it is. There is a problem even without a Cosmological Constant, as such an enormous energy would give the universe a strong positive curvature -- contrary to observations of a flat universe.

I am aware that when all matter and heat radiation have been removed from a region of space, even classically there still remains a pattern of zero-point electromagnetic field energy throughout the space. Nevertheless, as an interesting exercise let us see how much of the discrepancy can be reduced if we assume that the vacuum zero-point energy is only associated with matter, *i.e.* it only exists in the vicinity of matter due to a rapid attenuation of fields. In contrast to the prevailing view, in my speculative hypothesis this energy would be limited to the region of space near matter and there would only be negligible vacuum zero-point energy in the vast regions of space where there is no matter. If the volume of matter in the universe is mainly nucleons, we can easily estimate the reduction:

$$\frac{V_{Universe}}{V_{TotalNucleons}} \sim \frac{10^{79}\,m^3}{\left(\frac{M_{Universe}}{M_{Nucleon}}\right)V_{Nucleon}} \sim \frac{10^{79}\,m^3}{\left(\frac{10^{53}\,kg}{10^{-27}\,kg}\right)10^{-45}\,m^3} = 10^{44}. \qquad (9)$$

Thus the vacuum zero-point energy can be reduced to between $10^{46-44} = 10^2$ and $10^{120-44} = 10^{76}$ times too large if this premise has merit. At the low end, the discrepancy is almost resolved, but at the high end it is still unacceptably large. The discrepancy can be further reduced by including the free space within nucleons which are made of quarks. This divergence between experience and theory is perhaps the most bewildering problem in physics, whose resolution may lead to far-reaching consequences of our view of nature.

## 6. General Relativity

### 6.1 Einstein's general relativity

Einstein's General Relativity[5] (EGR) is one of the most profound creations of the human mind. But, EGR is a non-linear theory that becomes highly non-linear inside a black hole, resulting in a singularity. EGR not only predicts black holes, but also predicts that the mass inside the black hole will inevitably shrink down to a point, resulting in infinities of mass density, energy density, etc. Such singularities and other more subtle effects are indicative that at least in this regime EGR may be inconsistent with nature. This is similar to the breakdown of continuum hydrodynamics when length scales comparable to atomic diameters are considered.

In EGR all fields, except the gravitational field, produce space-time curvature, since the gravitational field is but a consequence of the curvature of space-time. Einstein reasoned that since curvature produced gravity, curvature cannot change gravity, *i.e.* make more or less gravity. To Einstein this would be double counting. The argument becomes less clear when inverted: Gravity cannot change curvature. This prohibition in EGR ultimately leads to black holes, in which singularities are the most egregious difficulties.

**6.2 Yilmaz' general relativity**

Hüseyin Yilmaz[6] made an interesting variation of EGR which avoids not only the singularities but the black holes altogether. He assumed that gravitation field energy also produces curvature of space-time by adding a "gravitational stress-energy tensor" to Einstein's equations. Since this term is relatively small in the three major tests of EGR, Yilmaz' General Relativity (YGR) makes essentially the same predictions as EGR for the advance of the perihelion of mercury, gravitational red shift, and the bending of starlight (the least accurately measured of the three tests). EGR and YGR cannot both satisfy the equivalence principle of GR -- at least not in its strong form. The Misner, Thorne, and Wheeler tome[7] and other texts cover many competing theories of gravitation and point out their shortcomings, but they seem not to have discussed or referenced Yilmaz' general relativity.

It may not be obvious why including gravitational field energy as a source of space curvature eliminates black holes. Here is a simple intuitive way to understand this. The static field and the near field (induction field) of a time-varying gravitational field have negative energy. (This can be tricky since the electrostatic field energy is positive. The radiation field has positive energy for both fields.) Negative energy gives negative curvature tending to cancel the positive curvature due to mass. Instead of black holes, YGR has grey holes where the emitted light is greatly red-shifted. If neutron stars with mass much greater than 3 solar masses, or white dwarfs with mass much smaller than 1.4 solar masses were detected, this would favor YGR over EGR.

Despite much effort, no one has yet solved the two-body problem in Einstein's General Relativity. Yilmaz and his colleagues would say that they never will since it is their view that EGR is a one-body theory in which one body (*e.g.* the Sun) establishes a space-time curvature (field) which determines the motion of a test body (*e.g.* Mercury) that hardly perturbs the established field. YGR claims to be an N-body theory. Yilmaz says that if one takes the weak field limit of EGR and considers a many body problem like the perturbation effects of planetary orbits on each other, EGR doesn't work. He asserts that the weak field limit of YGR gives exactly the same results as Newtonian gravity including perturbation effects. Another point in favor of Yilmaz is that in particle physics, energy is ascribed to the gravitational field.

I have mixed emotions and biases to overcome with respect to the absence of black holes in both YGR and Dicke's polarizable vacuum (ether) theory of gravitation. I certainly like the absence of singularities in these theories and they intrigue me. Until recently, I never had any reason to doubt the soundness of EGR. I did worry about black hole "no hair" theorems that appeared easily vulnerable to violation by the presence of other bodies. Misner[8] has challenged Yilmaz' formulation with rebuttal[9] by Alley and Yilmaz. At this point there is only a fracas. Whether it will develop into a full-fledged battle or remain a minor skirmish remains to be seen. If properly and fairly conducted, this is a healthy thing for physics, for honest dissension is a major mode by which physics progresses and comes into closer agreement with nature.

With respect to the 43" per century advance of the perihelion of Mercury predicted by both EGR and YGR, a special tribute is owed to astronomers as well as an accolade to Newtonian gravity which applies for the ordinary gravitational fields encountered in our solar system. It is not well known that the total perihelion advance of Mercury's orbit around the Sun is 575" per century of which 532" is due to the other planets as calculated by astronomers using Newtonian gravity. Paul Gerber,[10] effectively calculated 41"/century advance of the perihelion of mercury seventeen years before EGR using a retarded gravitational potential rather than Newton's instantaneous propagation of gravity. He calculated the speed at which gravity is propagated and got $3.055 \times 10^8$ m/sec., a speed slightly greater than today's speed of light, $c = 2.998 \times 10^8$ m/sec. He would have gotten 42"/century if he had used c. I would love to scrutinize his analysis in detail, but my ability to read German has gotten too rusty. I would be grateful for an English translation.

**6.3 Dicke's General Relativity**

Robert H. Dicke[11] developed a polarizable vacuum (ether) representation of General Relativity in which there is no space-time curvature. It derives gravitation as a manifestation of electromagnetism and would thus be a giant step forward toward a unified field theory if it were correct. It considers the implications of a greater ether density in the vicinity of a gravitating body. In his theory a body falls toward regions of greater gravitational field because the charged particles of which the body is composed move to the region of space where the dielectric constant of the ether is greater. Although Dicke's general relativity (DGR) has an entirely different basis than that of Yilmaz, the metric that Dicke obtains is precisely the same as Yilmaz. Hence it also has no black holes, and makes the same predictions for the three major experimental tests of GR. For all three approaches -- EGR, YGR, and DGR -- a light wave generates twice the gravitational field per unit energy density giving twice the deflection of a light ray than expected from Newtonian gravity.

As with YGR, DGR seems to have been totally neglected by the major chroniclers of GR such as Misner, Thorne, and Wheeler, and other texts. However Dicke and his student Carl Brans' new theory fared better, and was considered a major challenge to EGR for a while. It was motivated to clearly comply with Mach's principle. In the Brans-Dicke[11] GR theory, masses throughout the universe generate a scalar potential field, in addition to Einstein's curvature of spacetime, which can influence the strength of the universal gravitational constant G in space or time since G is linked to the matter-energy distribution of the universe. Whenever the scalar field is high, G is low. Dicke led theoretical and experimental challenges to GR. At the same time that he was promoting BDGR, he helped with experiments that brought about its downfall, and bolstered EGR. It is not clear what bearing these experiments had on YGR.

### 6.4 Mach's Principle

Local determination of an inertial frame can be made by observing a stationary bucket of water with a flat surface versus a concave parabolic surface for a non-inertial rotating pail, or by a Foucault pendulum. (A bucket of superfluid helium has its own surprises.[13]) Global determination of an inertial frame can be made by observation of a frame relative to the distant fixed stars i.e. the rest of the universe. It is remarkable that both methods give the same result. Mach was the first to conclude that this agreement implies that the distant stars determine the local inertial frame of reference since the local frame cannot conceivably affect the distant stars. Mach had trouble publishing his principle in the scientific journals of his day, so he included it in his 1912 book, *The Science of Mechanics*.[14]

Einstein was impressed that the distant stars have an influence on the local properties of matter, and tried to incorporate Mach's principle into General Relativity (GR), but succeeded in only a limited sense. GR does include Mach's principle in some special cases.[15] But this comes with some problems of its own. In EGR it is hard to see how those distant stars can impose their choice of which frames are inertial and which accelerating with sufficient alacrity to affect rapidly changing frames, if the frame information is limited to propagate at the speed of light. Aside from this problem, with respect to the rotating bucket, EGR does suggest that if the bucket is at rest and the distant stars rotate around it, its surface will be parabolic as if it were rotating. The BDGR theory was an attempt to better incorporate Mach's principle than does EGR.

### 6.5 Black or Grey Hole Radiation

Occasionally theories become established without a good foundation of supporting experiments. They become part of the lore of physics and are difficult to displace by a competing theory. Until the old theory is proven wrong by experiment, new paradigms are considered wrong by definition since they predict results which differ from those of the established, hence "correct" theory. This has been the case for Hawking's seniority model of black hole radiation which has been theoretically but not experimentally established.

If YGR is correct and there are no black holes, this impinges more on Hawking's model of black hole radiation than it does on my field emission-like model of gravitational tunneling radiation.[16-19] Tunneling radiation would still be nearly the same between a very dense little grey hole and a second body.

Belinski[20], a noted authority in the field of general relativity, unequivocally concludes "the effect [Hawking radiation] does not exist." He argues against Hawking radiation due to the infinite frequency of wave modes at the black hole horizon, and that the effect is merely an artifact resulting from an inadequate treatment of singularities.

The fact that a derivation is invalid does not disprove the existence of the effect. Belinski probes deeper than this by setting up the problem with what in his terms are proper finite wave modes. He then concludes that no particle creation can occur. The Rabinowitz radiation tunneling model involves no infinite frequency wave modes, and particle-antiparticle pairs do not have to be created.[16-19]

Hawking takes the position that radiation does not originate from within a black hole, but comes from the vicinity outside it, only appearing to come from within. Hawking concludes that the information that entered the black hole can be forever lost. If Hawking radiation does not come from within the hole (but only appears to) then it does not really reflect what is inside as the hole evaporates away. This is further exacerbated in that for him the radiation is black body radiation, which loses information about its source.

On the other hand, time-reversal symmetry, energy conservation, classical and quantum physics are violated if the information is lost. Gravitational tunneling radiation[16 - 19] resolves this enigma, since it comes from within the hole and carries attenuated but undistorted information from within. Since it is a tunneling process and not an information voiding Planckian black body radiation distribution, it can carry information related to the formation of a BH, and avoid the information paradox associated with Hawking radiation.

Parikh and Wilczek[21] take a black hole radiation tunneling approach quite similar to mine, but didn't reference my work. For them, the second body which makes the potential barrier finite, originates artificially from the black hole as a spherical shell. This seems a bit contrived to get a result similar to Hawking's with radiation in all directions. In my approach, the second body originates from outside the black hole and results in beamed radiation. Perhaps they didn't know about my papers, so I sent them copies.

**7. Little Holes and the Incidence Rate Of Ball Lightning**

It took centuries before the orthodoxy of science accepted ball lightning (BL) -- an experienced not theoretical phenomenon. It is my hypothesis that BL is a manifestation of cosmic little holes (LH). What follows helps to make a case for this. I've purposely left out the designation "black holes" or "grey holes" in the title of this section. The consequences of my analysis remain unchanged whether the holes are black or grey, so I will just call them "holes." In either case, they are candidates for the missing "dark" mass of the universe, and candidates as the cause of ball lightning and contributors to the accelerated expansion of the universe.

I will focus here on new analysis related to a prediction of the incidence rate of ball lightning not covered in my earlier work. The continuity equation for mass flow of LH when there is a creation rate $S_c$ and a decay rate $S_d$ of mass per unit volume per unit time t is

$$\nabla \bullet (\rho \vec{v}) + \partial\rho/\partial t = S_c - S_d, \quad (10)$$

where $\rho$ is the LH mass density at a given point in the universe, $\vec{v}$ is the LH velocity, and $\rho \vec{v}$ is the LH flux density. In steady state, $\partial\rho/\partial t = 0$. Integrating eq. (10):

$$\begin{aligned} &\int (\rho \vec{v}) \bullet d\vec{A} = \int (S_c - S_d)\, dV_t \Rightarrow \\ &-\rho_{LH} v_{LH} A_{far} + \rho_{BL} v_{BL} A_E = (S_c - S_d) V_t \end{aligned} \quad (11)$$

where $\rho_{LH}$ is the mass density of LH at a distance far from the earth, typical of the average mass density of LH throughout the universe. $A_{far}$ is the cross-sectional area of a curvilinear flux tube of LH far from the earth, $A_E$ is the cross-sectional area of the tube where it ends at the earth, and $V_t$ is the volume of the curvilinear flux tube (cylinder). Since the LH were created during the big bang, at a large distance from the earth they should be in the cosmic rest frame. The velocity of our local group of galaxies with respect to the microwave background (cosmic rest frame),[22] $v_{LH} \sim 6.2 \times 10^5$ m/sec is a reasonable velocity for LH with respect to the earth.

Because $v_{LH}$ is high and LH radiate little until they are near other masses, $S_c$ can be neglected because of negligible decay of large black holes into LH in the volume $V_t$. Similarly, $S_d$ may be expected to be small until LH are in the vicinity of the earth where most of their evaporation, before they are repelled away, is in a volume of the atmosphere ~ $A_E h$, where $A_E$ is the cross-sectional area of the earth, and h is a characteristic height above the earth. At this point it is helpful to convert to number density $\rho_L$ and $\rho_B$, of LH and ball lightning respectively. The number density decay rate is $r_B A_E h/t$, where t < ~ year is the dwell-time of LH near the earth. Thus equation (11) yields

$$\rho_B = \rho_L \left[ \frac{v_{LH}}{v_{BL} + (h/\tau)} \right] \frac{A_{far}}{A_E}, \quad (12)$$

which implies that the ball lightning flux is

$$\rho_B v_{BL} = \rho_L v_{LH}\left[\frac{v_{BL}}{v_{BL} + (h / \tau)}\right]\frac{A_{far}}{A_E} \approx \rho_L v_{LH}\left(\frac{A_{far}}{A_E}\right), (13)$$

where in most cases $h / t << v_{BL}$.

At large velocities, LH that do not slow down appreciably due to their large mass or angle of approach, either do not produce sufficient ionization to be visible or do not spend sufficient time in the atmosphere to be observed. In the Rabinowitz model,[16,18] those LH that reach the earth's atmosphere and are small enough to have sufficient radiation reaction force to slow them down to the range of $10^{-2}$ to $10^{2}$ m / sec, with a typical value $v_{BL} \sim 1$ m / sec, manifest themselves as BL. So equation (13) implies that the ball lightning current in the atmosphere ≈ the LH current far away. We can thus give a range for the BL flux density

$$\rho_L v_{LH} < \rho_B v_{BL} < \rho_L v_{LH}\left(\frac{A_{far}}{A_E}\right). \qquad (14)$$

The distribution of LH masses is not known. Assuming that LH comprise all of the dark matter, i. e. 95 % of the mass of the universe[16-18] of which there is a percentage p of LH of average mass $\overline{M}_{LH} \sim 10^{-3}$ kg:

$$\rho_L \sim \frac{p\left(0.95 M_{univ} / \overline{M}_{LH}\right)}{V_{univ}}. \qquad (15)$$

For $M_{univ} \sim 10^{53}$ kg, $V_{univ} \sim 10^{79}$ m$^3$ (radius of 15 x$10^9$ light-year = 1.4 x $10^{26}$ m), and p ~ 10 %, $\rho_L \sim 10^{-24}$ LH / m$^3$ . Thus from equations (12) and (13) my model predicts that the incidence rate of BL is roughly in the range $10^{-12}$ km$^{-2}$ sec$^{-1}$ to >~ $10^{-8}$ km$^{-2}$ sec$^{-1}$ for $A_{far} / A_E > \sim 10^4$. (Even if p = 100%, $10^{-11}$ km$^{-2}$ sec$^{-1}$ is well below the signal level of existing detectors.) This rate is in accord with the estimates of Barry and Singer[23] of 3 x $10^{-11}$ km$^{-2}$ sec$^{-1}$, and of Smirnov[24] of 6.4 x $10^{-8}$ km$^{-2}$ sec$^{-1}$ to $10^{-6}$ km$^{-2}$ sec$^{-1}$.

## 8. The Quantum and Classical Aharonov-Bohm and Similar Effects

### 8.1 Aharonov-Bohm effect

The question of which is more fundamental, force or potential energy -- or equivalently, field or potential-- is central to whether the laws of physics agree with the laws of nature. This is so even if it is decided that they are equal in importance. In Newtonian classical physics (NCP), force (*vis motrix* in Newton's Principia, 1686) and kinetic energy (*vis viva* in Leibnitz' Acta erud., 1695) are two of the foremost concepts. Although classically the concept of potential energy is a secondary concept for obtaining force and fields, its importance in the principle of the conservation of total energy cannot be overlooked. In quantum physics (QP), potential and kinetic energies are the primary concepts, with force hardly playing a role at all. One might incorrectly conclude: "No wonder then that effects were discovered early in the development of quantum physics that did not depend on forces at all." This was not the case, leaving aside the wave-particle duality, which is a horse of another color.[25]

It was not until 1959, some thirty-three years after the advent of quantum physics that Yakir Aharonov and David Bohm presented the first QP effect that was a radical departure from NCP. Aharonov and Bohm described gedanken electrostatic and magnetostatic cases in which physically measurable effects occur where no forces act.[26] This is now known as the Aharonov-Bohm effect.

In the magnetic case, an electron beam is sent around both sides of a long shielded solenoid or toroid so that the electron paths encounter no magnetic field and hence no magnetic force. They encounter a magnetic vector potential, which enters into the electron canonical momentum producing a phase shift of the electron wavefunction, and hence QP interference. If the electrons go through a double slit and screen apparatus[25] the shielded magnetic field shifts the interference pattern periodically as a function of the flux quantum in the shielded region, and hence Planck's constant. This was confirmed experimentally and considered a triumph for QP, and appears not to have been seriously challenged for forty-one years.

I read the A-B paper as a graduate student when it first came out in 1959. I understood it. It impressed me with the power of quantum physics. I did try to think of classical electrodynamic ways to explain the effect. Lilienfeld transition radiation and simpler effects entered my mind, but I did not pursue them. To Feynman the wave-particle duality as embodied in the two-slit experiment contains the mystery of QP.[25] To me the magic of QP was embodied in the A-B effect. (In my black hole analysis, I showed that there is an A-B effect for gravitational tunneling radiation.[17])

To my surprise and delight, I recently came across two papers by Timothy Boyer in which he reasons that the A-B effect can be understood completely classically.[27,28] [His approach is more direct than the one taken by Cohn and Rabinowitz[29] in developing a classical analog to quantum tunneling, which had also been thought to be completely QP.] First he points out that there has been no real experimental confirmation of the A-B effect. The periodic phase shift of a two-slit interference pattern due to a shielded magnetic field has indeed been confirmed. However, no experiment has shown that there are no forces on the electrons, that the electrons do not accelerate, and that the electrons on the two sides of a solenoid (or toroid) are not relatively displaced. Boyer then goes on to propose a classical mechanism that makes sense.

The charged particle (electron) induces a field in the conductor (shield or electromagnet, it doesn't seem to matter), and this field acts back on the charged particle producing a force which speeds up the particle as it approaches and then slows the particle as it recedes, so that it time averages to 0. This sequence is reversed on the other side of the magnetic source giving interference. The displaced charge in the shield (or solenoid windings) affects the current in the solenoid, and hence the center-of-energy of the solenoid field.

Boyer may not have tied all the loose ends together yet, but it is a promising start. With two thick superconducting shields -- separated by vacuum -- surrounding a toroid, it is not clear how the electron could classically sense the toroid's magnetic field, much less how Planck's constant h could enter into CP via the shielded magnetic flux. His explanation in terms of forces brings up problems with relativity; however, I think he is safe in considering the $v \ll c$ limit. Nevertheless his is the first real challenge to the A-B paradigm. At the very least, he has pointed out the way for the needed definitive experiments to decide whether or not the A-B effect can only be understood by QP.

**8.2 Aharonov-Casher effect**

Aharonov-Casher (A-C) in 1984 discovered an analog of the A-B effect in which the electrons are replaced by magnetic dipoles such as neutrons, and the shielded magnetic flux is replaced by a shielded line charge. Although it is considered to be solely in the domain of QP by the orthodox physics community, Boyer previously proposed a classical interpretation of the A-C phase shift effect.[27,28]

**8.3 Berry's geometric phase**

In 1984, the same year as the A-C effect, Michael V. Berry theoretically discovered that when an evolving quantum system returns to its original state, it has a memory of its motion in the geometric phase of its wavefunction. There are both quantum and classical examples of Berry's geometric phase (BGP), but as far as I know no one has yet challenged the QP case with a CP explanation. It is noteworthy that in 1992 Aharonov and Stern[30] did the QP analog of Boyer's[27,28] CP in analyzing BGP in terms of Lorentz-type and electric-type forces to show that BGP is analogous to the A-B effect.

**8.4 Winterberg's aether**

Friedwardt Winterberg developed an aether model using positive and negative Planck masses which he claims can lead to the results of quantum physics and relativity, as well as explaining the vacuum zero-point energy anomaly discussed in Section 5. He asserts that there is a close relationship between the Sagnac and A-B effects, which can be understood from his model.[31] For him, the A-B effect results from a cancellation of the kinetic energies of positive and negative mass vortices, much the same as the vanishing of the sum of positive and negative energies leading to ~ zero for the zero-point vacuum energy. Winterberg's theory is based upon the interesting concept that by assuming a different fundamental structure, NCP is adequate to obtain QP and EGR.

**9. Universe(s) According To Poe**

Over the ages there have been many differing views of the universe varying from finite to infinite in extent and / or time. In ancient eastern cosmology, not only the world, but the entire universe is believed to recycle itself in never-ending cycles of birth, growth, death, and rebirth from its own ashes, much like a phoenix. When it comes to imagining the universe, Edgar Allan Poe at the age of 39 (one year before his untimely death) was well ahead of his time. In his prescient science fiction writings he spoke of separate and distinct universes, each with its own set of laws -- much as only recently imagined by modern physics.[32] He writes:

> " there does exist a *limitless* succession of Universes, more or less similar to that of which we have cognizance ... it is abundantly clear that ... they have no portion in our laws. ... Each exists apart and independently ..."

In the preface to this book, Poe says that his book is for "the dreamers and those who put faith in dreams as the only realities." Poe was not only a dreamer, he was very well-read and knew the science of his day. In fact his book is dedicated to the eminent scientist Alexander von Humboldt.

Poe even had an interesting solution for what is known as Olber's paradox: If the universe is infinite and uniformly filled with stars, why isn't the night sky bright rather than dark? Poe reasoned that even if the universe were infinite in size, if the time since it came into existence were finite, there wasn't enough time for the light from very distant stars to have reached us to brighten the night sky.

## 10. CONCLUSION

The nature of reality, the fate of the cosmos, and the ultimate laws of the universe have been pondered for millennia. As insightful and far-sighted as past ages have been, our age clearly stands out as the era of profound discovery. We can say this not only because the vast majority of the discoveries have been made in our epoch, and not only because more than 98% of all the scientists that have ever lived are alive now. It is primarily because we have accumulated a critical mass of knowledge, thanks to the diligent efforts of scientists for hundreds of years. Thanks to these tireless servants of humanity, physics now probes the tiniest of sub-atomic particles and reaches out across the vast expanses of the universe.

Yet physics is a human enterprise fraught with human foibles. The laws of physics have far from perfect agreement with the laws of nature not only because humans are far from perfect, but as pointed out in the Introduction, possibly for reasons intrinsic to any mathematical or logical system. Nevertheless, because physics strives for a close correspondence with nature, the old laws are generally replaced by new laws that retain the old as limited cases. When we think we understand the laws of nature, we can contemplate mastering the universe. At the least, we can make the world a better place to live in. But it is a two-edged sword since great power for good is also great power for evil.

Physics is about modeling nature and reality. At its simplest level, physics only tries to describe nature. At deeper levels with more complex models, physics tries to predict and explain. Although modeling may be attempted at any given level of reality, it may not be able to go beyond certain levels or domains of validity for intrinsic reasons.

I don't subscribe to the Copenhagen view that reality is only what we can measure. I think reality is accessable to us at a deeper level than this. Some think of reality as direct experience such as experiments. But experiments depend on theory for interpretation. This is what is meant by saying experiments are theory-laden. In the beginning of my paper, I acknowledged that reality may not be the same to everyone. Physics tries in part to answer the question,"What is reality?" Nobel laureate, philosopher-poet Octavio Paz[33] said it beautifully :

> "Reality, everything we are, everything that envelops us, that sustains, and simultaneously devours and nourishes us, is richer and more changeable, more alive than all the ideas and systems that attempt to encompass it. ... Thus we do not truly know reality, but only the part of it we are able to reduce to language and concepts. What we call knowledge is knowing enough about a thing to be able to dominate and subdue it."

**Acknowledgment**

I wish to thank Hüseyin Yilmaz, Mark Davidson, and Arturo Alra Meuniot for valuable discussions. I am grateful to Friedwardt Winterberg for sending me a pre-publication copy of his intriguing book.

TABLE I. Wide Range of Experimental and Theoretical Transition Temperatures

| No. | Superconductor-fluid | $T_c^{exp}$ (K) | $T_c^{pred}$ (K) | $T_F^{2D}$ (K) | n ($10^{21}$/cm$^3$) | δ(Å) | $T_F^{3D}$ (K) |
|---|---|---|---|---|---|---|---|
| 1 | $^3He$ in $^4He$ | ? | $10^{-6}$ -$10^{-5}$ | - | - | - | - |
| 2 | H | ? | 30 x $10^{-6}$ | | $10^{-7}$ | | |
| 3 | $^3He$ | 0.0025 | 0.0028 | - | 23.6 | - | - |
| 4 | $^4He$ | 2.17 | 3 | - | 22 | - | - |
| 5 | $UPt_3$ | 0.53 | 0.68, 0.71, 0.74 | - | - | - | 124±5 |
| 6 | $[TMTSF]_2ClO_4$ | 1.2 | 2.31, 3.91, 6.35 | 72±24 | 0.38±0.17 | 13.275 | - |
| 7 | $TaS_2(Py)_{1/2}$ | 3.4 | 2.61, 3.60, 4.92 | 1710±360 | 12±3.6 | 12.02 | - |
| 8 | $PbMo_6S_8$ (Chevrel) | 12 | 7.38, 9.54, 11.68 | - | - | - | 1640±370 |
| 9 | κ-$[BEDT\text{-}TTF]_2Cu[NCS]_2$ | 10.5 | 7.26, 10.8, 15.3 | 213±57 | 0.31±0.09 | 15.24 | - |
| 10 | $BaPb_{0.75}Bi_{0.25}O_3$ | 11 | 11.4, 12.4, 13.4 | - | - | - | 242±20 |
| 11 | $Ba_{0.6}K_{0.4}BiO_3$ | 32 | 55.9, 62.1, 68.3 | - | - | - | 1210±120 |
| 12 | $La_{1.9}Sr_{0.1}CuO_4$ | 33 | 45.8, 50.8, 55.8 | 510±50 | ~0.5 | 6.61 | - |
| 13 | $La_{1.875}Sr_{0.125}CuO_4$ | 36 | 33.4, 56.2, 61.7 | 710±7 | ~1.0 | 6.61 | - |
| 14 | $La_{1.85}Sr_{0.15}CuO_4$ | 39 | 43.0, 49.7, 57.3 | 1090±100 | 5.2 ± 0.8 | 6.62 | - |
| 15 | $YBa_2Cu_3O_{6.67}$ | 60 | 49.8, 61.3, 79.4 | 710±70 | 1.1 ± 0.4 | 5.87 | - |
| 16 | $YBa_2Cu_4O_8$ | 80 | 64.3, 74.1, 87.8 | 1360±140 | 2.8 ± 0.44 | 6.81 | - |
| 17 | $YBa_2Cu_3O_7$ | 92 | 71.8, 79.9, 89.9 | 2290±100 | 16.9 ± 3.4 | 5.84 | - |
| 18 | $HoBa_2Cu_4O_8$ | 80 | 131, 145, 159 | 2070±200 | ~1.3 | 6.82 | - |
| 19 | $Bi_2Sr_2CaCu_2O_8$ | 89 | 37.6, 43.2, 77.1 | 970±390 | 3.5 ± 1.8 | 7.73 | - |
| 20 | $(Tl_{0.5}Pb_{0.5})Sr_2CaCu_2O_7$ | 80 | 75.5, 89.6, 103 | 1440±140 | 2.8 ± 0.5 | 6.05 | - |
| 21 | $Tl_2Ba_2CaCu_2O_8$ | 99 | 41.0, 49.6, 61.2 | 1180±110 | 4.9 ± 1.5 | 7.33 | - |

Where three numbers are shown for $T_c^{pred}$, these are the minimum, the mean, and the maximum predicted transition temperatures obtainable from the data as explained in ref. 1. The mean value does not always lie symmetrically between the minimum and maximum values.

TABLE II. Comparing Experimental and Theoretical Values of Pressure-Induced $T_c$

| Compound | $T_c^{exp}$ (K) | $T_{cP}^{exp}$ (K) | $T_{cP}^{theo}$ (K) | $\kappa$ $10^{-3}$ $(GPa)^{-1}$ | $\Delta P$ (GPa) |
|---|---|---|---|---|---|
| 1. $HgBa_2Ca_2Cu_3O_{8+\delta}$ | 135 | 164 | 166 | 10 | 31 |
| 2. $HgBa_2CaCa_2O_{8+\delta}$ | 128 | 154 | 157 | 10.6 | 29 |
| 3. $HgBa_2CuO_{8+\delta}$ | 94 | 118 | 113 | 11.4 | 24 |
| 4. $Bi_{1.68}Pb_{0.32}Ca_{1.85}Sr_{1.75}Cu_{2.65}O_{10}$ | 111 | 119 | 120 | 14.7 | 8 |
| 5. $Tl_2Ba_2CaCu_2O_8$ | 109 | 119 | 111 | 7.3 | 3 |
| 6. $YBa_2Cu_3O_{7-\delta}$ | 92 | 95 | 94.1 | 6.7 | 5 |
| 7. $YBa_2Cu_3O_{7-\delta}$ | 90 | 93 | 93.7 | 6.7 | 9 |
| 8. $Y_{0.9}Ca_{0.1}Ba_2Cu_4O_8$ | 90 | 98 | 92.7 | ≈ 9 | 5 |
| 9. $YBa_2Cu_4O_8$ | 80 | 101 | 83.2 | 8.5 | 7 |
| 10. $YBa_2Cu_4O_8$ | 80 | 106 | 84.7 | 8.5 | 10 |
| 11. $CaBaLaCu_3O_{6.85}$ | 62 | 86 | 64.8 | 8.2 | 8 |
| 12. $La_{1.85}Sr_{0.15}CuO_4$ | 36 | 41 | 36.8 | 6.9 | 4.5 |
| 13. $La_{1.85}Sr_{0.15}CuO_4$ | 36 | 41 | 36.7 | 6.79 | 4 |
| 14. $La_{1.8}Ba_{0.2}CuO_4$ | 36 | 44 | 36.1 | ≈ 5.3 | 1.1 |
| 15. $Nd_{1.32}Sr_{0.41}Ce_{0.27}CuO_{3.96}$ | 24 | 46 | >24.9 | 7.01 | >8 |

a) This is a compilation of superconductors for which sufficient data could be obtained to calculate the maximum pressure-induced transition temperature $T_{cP}^{theo}$ from the experimental atmospheric pressure transition temperature $T_c^{exp}$, in the pressure excursion ΔP, using only the compressibility κ as explained in ref.2. The calculated values $T_{cP}^{theo}$ are compared with the experimental values $T_{cP}^{exp}$.